\documentclass[twocolumn,showpacs,aps,prl,letterpaper]{revtex4}
\usepackage{graphicx}
\usepackage{dcolumn}
\usepackage{amsmath}
\usepackage{epsfig}
\usepackage{hyperref}

\long\def\inst#1{\par\nobreak\kern 4pt\nobreak
    {\itshape #1}\par\vskip 10pt plus 3pt minus 3pt}
\RequirePackage{xspace}
\usepackage{relsize}

\begin{document}

\title{\large \bfseries \boldmath Heavy-Light Mesons In A Relativistic Model}
\author{Jing-Bin Liu}\email{liujingbin077@mail.nankai.edu.cn}
\author{Mao-Zhi Yang}\email{yangmz@nankai.edu.cn}
\affiliation{School of Physics, Nankai University, Tianjin 300071,
P.R. China}

\date{May 13, 2015}

\begin{abstract}
We study the heavy-light mesons in a relativistic model, which is derived from the Bethe-Salpeter equation by applying the Foldy-Wouthuysen transformation on the heavy quark. The kernel we choose is based on scalar confining and vector Coulomb potentials, the transverse interaction of the gluon exchange is also taken into account in this model. The spectra and wave functions of $D$, $D_s$, $B$, $B_s$ meson states are obtained. The spectra are calculated up to the order of $1/m_Q$, and wave functions are treated to leading order.
\end{abstract}

\pacs{12.39.Pn, 14.40.Lb, 14.40.Nd}

\maketitle

\section*{I Introduction}

Recently, great progress has been made in the measurement of the spectrum of heavy-light quark-antiquark system, especially in the charm sector. Several new excited states were established in  both $D$ and $D_s$  mesons \cite{Ds2632,Ds2860,Ds2700,Ds2009,LHCbspin13b,LHCbspin13a,D2010,LHCb}. Many theoretical models were propounded to identify the new discovered states and explain their properties \cite{CF1,LJM,BB,CFGN,SSC,LY2}. The spectroscopy provides a powerful test of the theoretical predictions based on the quark model in the standard model.

Heavy-light quark-antiquark system $Q\bar{q}$ plays an important role in understanding the strong interactions between quark and antiquark. Double-heavy meson sector is rather mature, and nonrelativistic potential models have been proven extremely successful for the description for heavy quarkonia \cite{LSG}. As for heavy-light meson system, one needs a model that can include the relativistic effects of the light quark.

In order to describe the relativistic effects in the bound states, we resort to the Bethe-Salpeter equation. Many studies of mesons  were carried out in the Bethe-Salpeter approach \cite{YK,WWC,FKW1,FKW2,HP}. It is very difficult to solve the Bethe-Salpeter equation, especially when we consider states with large total angular momentum number $J$. We can reduce the Bethe-Salpeter equation to an approximate but easier form, so as to study mesons systematically. We apply the Foldy-Wouthuysen transformation on the heavy quark and expand the interaction terms to order $1/m_Q$. The instantaneous approximation can only contribute to corrections of order $(1/m_Q)^2$ if one considers the heavy quark on shell as in the heavy quark effective theory (HQET). In this paper we take the instantaneous approximation for the Bethe-Salpeter equation.

One can bring the instantaneous Bethe-Salpeter equation into the Breit form by taking further approximation. The Breit interaction is widely used to study quark-antiquark systems as an nonrelativistic or semirelativistic model \cite{MKM,TM,MM}. While in this paper, we take the instantaneous Bethe-Salpeter equation in its original form, from which one can derive a more reliable model that preserves the relativistic effects of the light quark.

We treat the equation in the $1/m_Q$ expansion to the leading order. The parameters in the equations are determined by fitting the spectra of the heavy-light meson states presented in Particle Data Group (PDG) \cite{PDG}. The identifications of the newly observed heavy-light meson states are given based on the theoretical prediction.

The organization of this paper is as follows. Section II is for the relativistic model and the effective Hamiltonian of the heavy-light quark-antiquark system. In section III, we solve the relativistic wave equation. Section IV is for the numerical result and discussion. In section V we have a brief summary.

\section*{II The model}
The Bethe-Salpeter Equation for the quark-antiquark system can be written as \cite{BS}:
\begin{equation}
(p\!\!\!/_1-m_1)\chi(p)(p\!\!\!/_2+m_2)=\int\!\frac{d^4p^\prime}{(2\pi)^4}\overline K(p,p^\prime,P)\chi(p^\prime) ,   \label{e1}
\end{equation}
where the total momentum $P$ (in the center-of-mass system, $P=(E,\boldsymbol{0})$) and the relative momentum $p$ are defined as:
\begin{eqnarray}
P&=&p_1+p_2,\\
p&=&\alpha_1p_2-\alpha_2p_1
\end{eqnarray}
with
\begin{equation}
\alpha_i=\frac{m_i}{m_1+m_2}, i=1,2.
\end{equation}

It is necessary to assume some form for the kernel in order to reduce the Bethe-Salpeter equation. Here we take the kernel as the simplest form \cite{JZ},  which can be reduced to that used in Ref. \cite{GI}. The kernel can be written as:
\begin{equation}
\overline K(p,p^\prime,P)=\gamma^{(1)}\cdot\gamma^{(2)}V_v(k^2)+V_s(k^2),  \label{e2}
\end{equation}
where $k$ is the transferred four-momentum,
\begin{equation}
k=p-p^\prime.
\end{equation}
Inserting Eq. (\ref{e2}) into Eq. (\ref{e1}), the Bethe-Salpeter equation can be
written as:
\begin{eqnarray}
&&
(p\!\!\!/_1-m_1)\chi(p)(p\!\!\!/_2+m_2)=\nonumber\\
&&
\int\!\frac{d^4k}{(2\pi)^4}\left[V_v(k^2)\gamma^{(1)}\cdot
\chi(p+k)\gamma^{(2)}+V_s(k^2)\chi(p+k)\right]. \nonumber\\ \label{e3}
\end{eqnarray}
The above equation can be transformed to a more convenient form that is symmetric with respect to particles and antiparticles by using the charge conjugation transformation. We rewrite Eq. (\ref{e3}) as:
\begin{eqnarray}
&&
(p\!\!\!/_1-m_1)(p\!\!\!/_2+m_2)^T\chi(p)=\nonumber\\
&&
\int\!\frac{d^4k}{(2\pi)^4}\left[V_v(k^2)\gamma^{(1)}\cdot{\gamma^{(2)}}^T
\chi(p+k)+V_s(k^2)\chi(p+k)\right]. \nonumber\\\label{e4}
\end{eqnarray}
After applying the condition
\begin{equation}
\mathcal{C}{\gamma^\mu}^T\mathcal{C}^{-1}=-\gamma^\mu,
\end{equation}
we arrive at
\begin{eqnarray}
&&
(p\!\!\!/_1-m_1)(p\!\!\!/_2-m_2)\psi(p)=\nonumber\\
&&
\int\!\frac{d^4k}{(2\pi)^4}\left[V_v(k^2)\gamma^{(1)}\cdot{\gamma^{(2)}}
\psi(p+k)-V_s(k^2)\psi(p+k)\right], \nonumber\\\label{e5}
\end{eqnarray}
where $\psi(p)$ is the charge conjugated wave function, and the minus sign appeared before $V_s$ indicates that the confinement force is always attractive.

By taking the instantaneous approximation, we can perform the $p^0$ integration and decompose the instantaneous Bethe-Salpeter equation into four equations. Details of this derivation can be found in Refs. \cite{GR,CCW}.

The projection operators are defined as
\begin{eqnarray}
&&\Lambda^{(i)}_{\pm}=\frac{1}{2}(1{\pm}h_i),\\
&&h_i=\frac{H_i}{\omega_i}, \label{a2}\\
&&\omega_i=\sqrt{\boldsymbol{p}^2+m_i^2},\;\; i=1,2
\end{eqnarray}
with
\begin{eqnarray}
H_1(\boldsymbol{p})&=&-{\boldsymbol{\alpha}}^{(1)}\cdot\boldsymbol{p}+{\beta}^{(1)}m_1,\\
H_2(\boldsymbol{p})&=&{\boldsymbol{\alpha}}^{(2)}\cdot\boldsymbol{p}+{\beta}^{(2)}m_2.\label{e37}
\end{eqnarray}
Applying the projection operators, the four coupled equation can be written as
\begin{eqnarray}
(E-\omega_1-\omega_2)\phi_{++}(\boldsymbol{p})&=&-2{\pi}i \Gamma_{++}(\boldsymbol{p}),\label{e6}\\
(E+\omega_1+\omega_2)\phi_{--}(\boldsymbol{p})&=&2{\pi}i \Gamma_{--}(\boldsymbol{p}),\label{e7}\\
\phi_{+-}(\boldsymbol{p})&=&0, \label{e8} \\
\phi_{-+}(\boldsymbol{p})&=&0, \label{e9}
\end{eqnarray}
where
\begin{eqnarray}
&&\phi(\boldsymbol{p})=\int\!dp^0\psi(p^0,\boldsymbol{p}),\\
&&\Gamma(\boldsymbol{p})=\int\!\frac{d^3k}{(2\pi)^4}\gamma_0^{(1)}\gamma_0^{(2)}
\left[V_v(-\boldsymbol{k}^2)\gamma^{(1)}\cdot{\gamma^{(2)}}\right.\nonumber\\
         &&\left.-V_s(-\boldsymbol{k}^2)\right]\phi(\boldsymbol{p}+\boldsymbol{k}),\\
&&\phi_{\pm\pm}=\Lambda^{(1)}_\pm\Lambda^{(2)}_\pm\phi,\;\; \Gamma_{\pm\pm}=\Lambda^{(1)}_\pm\Lambda^{(2)}_\pm\Gamma.
\end{eqnarray}

One can combine the four coupled equations and obtain the equation of $\phi$ instead of
its projective components $\phi_{\pm\pm}$ as in Ref. \cite{ES}, that is
\begin{equation}
(E-H_1(\boldsymbol{p})-H_2(\boldsymbol{p}))\phi(\boldsymbol{p})=-2{\pi}i \Lambda(\boldsymbol{p}) \Gamma(\boldsymbol{p}),\\ \label{e10}
\end{equation}
where
\begin{eqnarray}
\Lambda(\boldsymbol{p})&=&\Lambda^{(1)}_+\Lambda^{(2)}_+-\Lambda^{(1)}_-\Lambda^{(2)}_-
\nonumber\\
&=&\frac{1}{2}(h_1(\boldsymbol{p})+h_2(\boldsymbol{p})).
\end{eqnarray}

If one makes the approximation
\begin{equation}
 \Lambda(\boldsymbol{p})\rightarrow 1, \label{e17}
\end{equation}
Eq. (\ref{e10}) can be bought into the Breit form. The Breit equation can be used as an nonrelativistic or semirelativistic approach to study the interactions of quarks. In order to take the relativistic corrections of the light quark into account, we keep $\Lambda(\boldsymbol{p})$ in its original form instead of taking it as 1.

Eq. (\ref{e10}) can be transformed into coordinate space according to
\begin{equation}
\phi(\boldsymbol{r})=\int\!\frac{d^3p}{(2\pi)^3}e^{i\boldsymbol{p}\cdot\boldsymbol{r}}\phi
(\boldsymbol{p}).
\end{equation}
The coordinate form of Eq. (\ref{e10}) is
\begin{equation}
\left(H_1+H_2+\frac{1}{2}(h_1+h_2)U\right)\phi(\boldsymbol{r})=E\phi(\boldsymbol{r})
\label{e11}
\end{equation}
with
\begin{eqnarray}
U(\boldsymbol{r})&=&U_1(\boldsymbol{r})+U_2(\boldsymbol{r}),\\
U_1(\boldsymbol{r})&=&V_v(r)+\beta^{(1)}\beta^{(2)}V_s(r),\\
U_2(\boldsymbol{r})&=&-\frac{1}{2}[{\boldsymbol{\alpha}}^{(1)}\cdot{\boldsymbol{\alpha}}^
{(2)}+({\boldsymbol{\alpha}}^{(1)}\cdot\hat{\boldsymbol{r}})
({\boldsymbol{\alpha}}^{(2)}\cdot\hat{\boldsymbol{r}})]V_v(r),\nonumber\\
\end{eqnarray}
where $V_v(r)$ and $V_s(r)$ are the Fourier transformations of $V_v(-\boldsymbol{k}^2)$ and
$V_s(-\boldsymbol{k}^2)$, respectively.

The wave function $\phi(\boldsymbol{r})$ has two spinor indices, i.e. $4\times4=16$ components, it is difficult to solve an eigenequation with so many components. For the heavy-light system, one can reduce the eigenequation by applying the nonrelativistic approximation of the heavy quark $Q$ and expand the interactions in the order of $1/m_Q$.

In an elegant and systematic way the reduction can be achieved by using the Foldy-Wouthuysen transformation. Here we designate the heavy quark $Q$ and the light antiquark $\bar{q}$ by the index 1 and 2 in Eq. (\ref{e11}), respectively. Thus the transformation is performed on the heavy quark with the superscript ``1". We reduce the Hamiltonian in the Dirac representation, where
\begin{eqnarray}
\boldsymbol{\alpha}=\left( \begin{array}{cc} & \boldsymbol{\sigma}\\\boldsymbol{\sigma}& \end{array}\right ),
\beta=\left( \begin{array}{cc} 1& \\ & -1 \end{array}\right ),
\boldsymbol{\Sigma}=\left( \begin{array}{cc} \boldsymbol{\sigma}& \\& \boldsymbol{\sigma}\end{array}\right ).
\end{eqnarray}

If the original Hamiltonian is written in the form
\begin{equation}
H=\beta m+\mathcal{E}+\mathcal{O},
\end{equation}
where $\mathcal{O}$ is the ``odd" operator, typical examples are the matrices $\boldsymbol{\alpha}$ and $\boldsymbol{\gamma}$, while $\mathcal{E}$ is the ``even" operator, examples of this class of operators are 1, $\beta$ and $\boldsymbol{\Sigma}$. According to the Foldy-Wouthuysen transformation, the transformed Hamiltonian reads
\begin{eqnarray}
\tilde{H}&=&U_F^{-1}HU_F\nonumber\\
&=&\beta m+\mathcal{E}+\frac{\beta}{2m}\mathcal{O}^2+\frac{1}{8m^2}[ [ \mathcal{O},\mathcal{E} ],
\mathcal{O} ]-\frac{\beta}{8m^3}\mathcal{O}^4+\cdots \nonumber\\\label{e20}
\end{eqnarray}

The Hamiltonian in Eq. (\ref{e11}) is
\begin{equation}
H=H_1+H_2+\frac{1}{2}(h_1+h_2)U.
\label{e12}
\end{equation}
If we perform the Foldy-Wouthuysen transformation directly on Eq. (\ref{e12}), the result can be awkward. What we want to achieve is to obtain a reduced and simple form of the Hamiltonian for the heavy-light quark-antiquark system. Here we show that it is possible. It should be noticed that Eq. (\ref{e11}) is not equivalent to the instantaneous Bethe-Salpeter equation, i.e. the
four coupled equations Eqs. (\ref{e6})$\sim$(\ref{e9}), as pointed out in Refs. \cite{CCL,CC}. We find that our goal can be achieved by employing the constraint of Eqs. (\ref{e8}) and (\ref{e9}) which are dropped in Eq. (\ref{e11}).

Subtracting Eq. (\ref{e9}) from Eq. (\ref{e8}), we get
\begin{equation}
(h_1-h_2)\phi=0. \label{e15}
\end{equation}

We write Eq. (\ref{e11}) together with Eq. (\ref{e15}) as:
\begin{eqnarray}
\left(H_1+H_2+\frac{1}{2}(h_1+h_2)U-E\right)\phi(\boldsymbol{r})&=&0,\label{e100}\\
(h_1-h_2)\phi(\boldsymbol{r})&=&0. \label{e101}
\end{eqnarray}

It is easy to verify that the above two equations are equivalent to the instantaneous Bethe-Salpeter equation, i.e. the four coupled equations Eqs. (\ref{e6})$\sim$(\ref{e9}). Here we expect to obtain one single equation that is equivalent to the above two equations by combining them.

Eq. (\ref{e101}) is equivalent to
\begin{equation}
\phi=\frac{1}{2}(h_1+h_2)\varphi.\label{e102}
\end{equation}
If the above equation is true, then
\begin{equation}
(h_1-h_2)\phi=\frac{1}{2}(h_1-h_2)(h_1+h_2)\varphi=0,
\end{equation}
on the other hand, if Eq. (\ref{e101}) is true, we take $\varphi=h_1\phi$, then
\begin{equation}
\frac{1}{2}(h_1+h_2)\varphi=\frac{1}{2}(1+h_1h_2)\phi=\phi.
\end{equation}

Applying the equivalent form of Eq.(\ref{e101}), Eqs.(\ref{e100}) can be transformed to a new equivalent equation
\begin{equation}
\left(H_1+H_2+\frac{1}{2}(h_1+h_2)U-E\right)\frac{1}{2}(h_1+h_2)\varphi=0,\label{e103}
\end{equation}
it is easy to verify that the above equation is equivalent to the two coupled equations (\ref{e100}) and (\ref{e101}). Furthermore, it is equivalent to
\begin{eqnarray}
&&\left(\frac{1}{2}(1+h_1h_2)(\omega_1+\omega_2)+\frac{1}{2}(h_1+h_2)U\frac{1}{2}(h_1+h_2)\right.\nonumber\\
&&\left.-\frac{1}{2}(h_1+h_2)E\right)\varphi=0,\label{e104}
\end{eqnarray}

Noticing the relations
\begin{eqnarray}
(1+h_1h_2)\frac{1}{2}(h_1+h_2)&=&h_1+h_2,\label{e105}\\
(1+h_1h_2)h_1&=&h_1+h_2,\label{e106}\\
(1+h_1h_2)h_2&=&h_1+h_2,\label{e107}
\end{eqnarray}
we can have two different equivalent forms of Eq. (\ref{e104}), they are
\begin{eqnarray}
&&\frac{1}{2}(1+h_1h_2)\left(\omega_1+\omega_2+\frac{1}{2}(h_1+h_2)U\frac{1}{2}(h_1+h_2)\right.\nonumber\\
&&\left.-\frac{1}{2}(h_1+h_2)E\right)\varphi=0,\label{e108}
\end{eqnarray}
and
\begin{eqnarray}
&&\left(\omega_1+\omega_2+\frac{1}{2}(h_1+h_2)U\frac{1}{2}(h_1+h_2)\right.\nonumber\\
&&\left.-\frac{1}{2}(h_1+h_2)E\right)\frac{1}{2}(1+h_1h_2)\varphi=0.\label{e109}
\end{eqnarray}

Here we guess a form that is equivalent to Eq. ({\ref{e108}) and Eq. (\ref{e109}}) by employing the common part of them, that is
\begin{eqnarray}
&&\left(\omega_1+\omega_2+\frac{1}{2}(h_1+h_2)U\frac{1}{2}(h_1+h_2)\right.\nonumber\\
&&\left.-\frac{1}{2}(h_1+h_2)E\right)\psi=0.\label{e110}
\end{eqnarray}

Left multiplying the above equation by $\frac{1}{2}(1+h_1h_2)$, we get the form of Eq. (\ref{e108}). On the other hand, we can obtain the form of the above equation  from Eq. (\ref{e109}) if we take $\psi=\frac{1}{2}(1+h_1h_2)\varphi$.
That is to say, Eq. (\ref{e110}) is equivalent to Eq. (\ref{e108}) and Eq. (\ref{e109}), thus equivalent to
Eqs. (\ref{e100}) and (\ref{e101}), i.e. the instantaneous Bethe-Salpeter equation.

By using Eq.(\ref{e106}) and Eq.(\ref{e107}), we can obtain two other equivalent equations,
respectively.
\begin{eqnarray}
&&\left(\omega_1+\omega_2+\frac{1}{2}(h_1+h_2)U\frac{1}{2}(h_1+h_2)-h_1E\right)\psi=0,\label{e111}\nonumber\\
\\
&&\left(\omega_1+\omega_2+\frac{1}{2}(h_1+h_2)U\frac{1}{2}(h_1+h_2)-h_2E\right)\psi=0.\label{e112}\nonumber\\
\end{eqnarray}

In the foregoing paragraphs, equivalent equations of the instantaneous Bethe-Salpeter equation are obtained. In Appendix A, the reduction of the instantaneous Bethe-Salpeter equation for
nonrelativistic systems is discussed.  We find that the reductions of the Hamiltonian with the approximation (\ref{e17}), i.e. the Breit interaction, and without the approximation (\ref{e17}), i.e. the choice in this paper,  do not differ until order $1/m_im_j$. While for the case of heavy-light system we focus in this paper, the reduction results of the two scheme can be essentially different.

Now we return to the reduction of the Hamiltonian for the heavy-light system.
Eq. (\ref{e111}) can be written as (the wave function in the equation is omitted):
\begin{eqnarray}
-h_1E+_\Delta\!\omega_1+\omega_2+\frac{1}{2}(h_1+h_2)U\frac{1}{2}(h_1+h_2)=-m_1,\nonumber\\\label{e201}
\end{eqnarray}
where
\begin{eqnarray}
_\Delta\omega_1=\omega_1-m_1,\label{e202}
\end{eqnarray}
now we consider performing the Foldy-Wouthuysen transformation on Eq. (\ref{e201}), the left of it can be rewritten as:
\begin{eqnarray}
&&-\beta^{(1)} E-E\left(\frac{m_1}{\omega_1}-1\right)\beta^{(1)}-E\left(-\frac{\boldsymbol{\alpha}^{(1)}\cdot\boldsymbol{p} }{\omega_1}\right)\nonumber\\
&&+_\Delta\!\omega_1+\omega_2+\frac{1}{2}(h_1+h_2)U\frac{1}{2}(h_1+h_2)\nonumber\\
&&=-\beta^{(1)} E+\mathcal{E}+\mathcal{O}, \label{e203}
\end{eqnarray}
the odd and even operators in the above equation are:
\begin{eqnarray}
\mathcal{E}&=&-E\left(\frac{m_1}{\omega_1}-1\right)\beta^{(1)}+_\Delta\!\omega_1+\omega_2
\nonumber\\
&+&\frac{1}{2}\left(\frac{m_1}{\omega_1}\beta^{(1)}+h_2\right)U_1\frac{1}{2}\left(\frac{m_1}{\omega_1}\beta^{(1)}+h_2\right)
\nonumber\\
&+&\frac{1}{2}\left(-\frac{\boldsymbol{\alpha}^{(1)}\cdot\boldsymbol{p} }{\omega_1}\right)U_1\frac{1}{2}\left(-\frac{\boldsymbol{\alpha}^{(1)}\cdot\boldsymbol{p} }{\omega_1}\right)
\nonumber\\
&+&\frac{1}{2}\left(-\frac{\boldsymbol{\alpha}^{(1)}\cdot\boldsymbol{p} }{\omega_1}\right)U_2\frac{1}{2}\left(\frac{m_1}{\omega_1}\beta^{(1)}+h_2\right)
+h.c.
\end{eqnarray}
and
\begin{eqnarray}
\mathcal{O}&=&-E\left(-\frac{\boldsymbol{\alpha}^{(1)}\cdot\boldsymbol{p} }{\omega_1}\right)\nonumber\\
&+&\frac{1}{2}\left(-\frac{\boldsymbol{\alpha}^{(1)}\cdot\boldsymbol{p} }{\omega_1}\right)U_1\frac{1}{2}\left(\frac{m_1}{\omega_1}\beta^{(1)}+h_2\right)
+h.c.
\nonumber\\
&+&\frac{1}{2}\left(\frac{m_1}{\omega_1}\beta^{(1)}+h_2\right)U_2\frac{1}{2}\left(\frac{m_1}{\omega_1}\beta^{(1)}+h_2\right)
\nonumber\\
&+&\frac{1}{2}\left(-\frac{\boldsymbol{\alpha}^{(1)}\cdot\boldsymbol{p} }{\omega_1}\right)U_2\frac{1}{2}\left(-\frac{\boldsymbol{\alpha}^{(1)}\cdot\boldsymbol{p} }{\omega_1}\right)
\end{eqnarray}
where $h.c.$ stands for Hermitian conjugate.

By using the new form of Eq. (\ref{e20})
\begin{eqnarray}
\tilde{H}
=-\beta^{(1)}E+\mathcal{E}-\frac{\beta^{(1)}}{2E}\mathcal{O}^2+\frac{1}{8E^2}[ [ \mathcal{O},\mathcal{E} ],
\mathcal{O} ]+\cdots \label{e210}
\end{eqnarray}
we can expand Eq. (\ref{e201}) to order $1/E$£º
\begin{eqnarray}
-\beta^{(1)}E+\tilde{H}_0+\tilde{H}_1+\tilde{H}_2=0,\label{e211}
\end{eqnarray}
where
\begin{eqnarray}
\tilde{H}_0&=&\omega_1+\omega_2
+\frac{1}{2}\left(\beta^{(1)}+h_2\right)U_1\frac{1}{2}\left(\beta^{(1)}+h_2\right),
\\
\tilde{H}_1&=&\frac{1}{2}\left(-\frac{\boldsymbol{\alpha}^{(1)}\cdot\boldsymbol{p} }{\omega_1}\right)U_2\frac{1}{2}\left(\beta^{(1)}+h_2\right)
+h.c.
\nonumber\\
&-&\frac{\beta^{(1)}}{2m_1}\{\alpha^{(1)}\cdot\boldsymbol{p},\;\frac{1}{2}(\beta^{(1)}+h_2)U_2\frac{1}{2}(\beta^{(1)}+h_2)\},
\nonumber\\
\\
\tilde{H}_2&=&-\frac{\beta^{(1)}}{2E}\left[\frac{1}{2}(\beta^{(1)}+h_2)U_2\frac{1}{2}(\beta^{(1)}+h_2)\right]^2, \label{e212}
\end{eqnarray}
The last term $\tilde{H}_2$ is of the order $V_v^2$. Comparing to the other terms, its correction is suppressed by the coupling constant, here we omit it. Considering the substitution $\beta^{(1)}\rightarrow1$ in the Foldy-Wouthuysen transformation, $\tilde{H}_1$ can be reduced to
\begin{eqnarray}
\tilde{H}_1&\Rightarrow&\frac{1}{2}\left(-\frac{\boldsymbol{\alpha}^{(1)}\cdot\boldsymbol{p} }{\omega_1}\right)U_2\frac{1}{2}\left(1+h_2\right)
+h.c.
\nonumber\\
&+&\frac{1}{2}(-1+h_2)\frac{1}{2}\left(-\frac{\boldsymbol{\alpha}^{(1)}\cdot\boldsymbol{p} }{m_1}\right)U_2\frac{1}{2}(1+h_2)+h.c.
\nonumber\\
&=&\frac{1}{2}(1+h_2)\frac{1}{2}\left(-\frac{\boldsymbol{\alpha}^{(1)}\cdot\boldsymbol{p} }{m_1}\right)U_2\frac{1}{2}(1+h_2)+h.c.
\nonumber\\
&=&\frac{1}{2}(1+h_2)\;\frac{1}{2}\{-\frac{{\boldsymbol{\alpha}}^{(1)}\cdot\boldsymbol{p}}{m_1}, U_2\}\;\frac{1}{2}(1+h_2),
\end{eqnarray}
then $\tilde{H}$ is transformed to£º
\begin{eqnarray}
\tilde{H}&=&\omega_1+\omega_2
+\frac{1}{2}\left(1+h_2\right)U_1\frac{1}{2}\left(1+h_2\right)\nonumber\\
&+&\frac{1}{2}(1+h_2)\;\frac{1}{2}\{-\frac{{\boldsymbol{\alpha}}^{(1)}\cdot\boldsymbol{p}}{m_1}, U_2\}\;\frac{1}{2}(1+h_2),
\label{e215}
\end{eqnarray}
and we obtain the final Hamiltonian
\begin{eqnarray}
H&=&H_0+H^\prime \label{e216}\\
H_0&=&\omega_1+\omega_2+\frac{1}{2}(1+h_2)(V_v+\beta^{(2)}V_s)\frac{1}{2}(1+h_2)\label{e217}\\
H^\prime&=&\frac{1}{2}(1+h_2)
\{\frac{ V_v}{4m_1}\left[({\boldsymbol{\alpha}}^
{(2)}+({\boldsymbol{\alpha}}^{(2)}\cdot\hat{\boldsymbol{r}})\hat{\boldsymbol{r}})
\cdot\boldsymbol{p}\right.\nonumber\\
&+&\left.i\boldsymbol{\sigma}^{(1)}\cdot
({\boldsymbol{\alpha}}^
{(2)}+({\boldsymbol{\alpha}}^{(2)}\cdot\hat{\boldsymbol{r}})\hat{\boldsymbol{r}})
\times\boldsymbol{p}\right]+h.c.\}
\frac{1}{2}(1+h_2)\nonumber\\
\label{e218}
\end{eqnarray}

The scalar and vector potentials we choose have the simple form
\begin{eqnarray}
V_s(r)=br+c,\\
V_v(r)=-\frac{4\alpha_s}{3r}.
\end{eqnarray}
The potentials are chosen to have a Coulombic behavior at short distance and a linear confining behavior at long distance.

\section*{III Solution of the Wave Equation}

In this section, we solve the eigenequation of $H_0$ and then deal with the perturbative corrections of $H^\prime$.

In order to generalize the algorithm which was used in Refs. \cite{ymz,LY1}, we prove a relation in the first place.

Let $\Omega(p)$ be the pseudo-differential operator function and $\Omega(k)$ be the normal function, here $p$ and $k$ stand for the modules of momentum operator $\boldsymbol{p}$ and momentum $\boldsymbol{k}$, respectively. Let $\psi(\boldsymbol{r})$  be a function of $\boldsymbol{r}$ that can be written as
\begin{equation}
\psi(\boldsymbol{r})=\phi(r)Y_{lm}(\hat{\boldsymbol{r}}),\label{e28}
\end{equation}
with the help of
\begin{equation}
\delta^3(\boldsymbol{r}-\boldsymbol{r}^\prime)=\int\frac{d^3k}{(2\pi)^3}e^{i\boldsymbol{k}\cdot
(\boldsymbol{r}-\boldsymbol{r}^\prime)},
\end{equation}
we have
\begin{eqnarray}
&&\Omega(p)\psi(\boldsymbol{r})=\Omega(p)\int d^3r^{\prime}\;\delta
^3(\boldsymbol{r}-\boldsymbol{r}^{\prime})\psi({\boldsymbol{r}}^{\prime})\nonumber\\
&&=\int d^3r^{\prime}\int \frac{d^3k}{(2\pi )^3}\Omega(k)
e^{i\boldsymbol{k}\cdot
(\boldsymbol{r}-\boldsymbol{r}^\prime)}\psi(\boldsymbol{r}^\prime).\label{e29}
\end{eqnarray}

The exponential factor $e^{i\boldsymbol{k}\cdot\boldsymbol{r}}$ can be decomposed into series of spherical harmonics
\begin{equation}
e^{i\boldsymbol{k}\cdot\boldsymbol{r}}=4\pi
\sum_{lm}i^lj_l(kr)Y^*_{lm}(\hat{\boldsymbol{k}})Y_{lm}(\hat{\boldsymbol{r}}),\label{e30}
\end{equation}
where $j_l$ is the $l$-th order spherical Bessel function, $Y_{lm}(\hat{\boldsymbol{r}})$ is the spherical harmonics, which satisfies the normalization condition
\begin{equation}
\int d\Omega
Y_{l_1m_1}^*(\hat{\boldsymbol{r}})Y_{l_2m_2}(\hat{\boldsymbol{r}})=\delta_{l_1l_2}\delta_{m_1m_2},
\end{equation}
with $\hat{\boldsymbol{r}}$ the unit vector along the direction of $\boldsymbol{r}$.

Inserting Eqs. (\ref{e28}) and (\ref{e30}) into Eq. (\ref{e29}), after simplification we arrive at
\begin{eqnarray}
&&\Omega(p)\phi(r)Y_{lm}(\hat{\boldsymbol{r}})\nonumber\\
&&=\frac{2}{\pi}\int dr^{\prime}{r^{\prime}}^2\int dk\,k^2\Omega(k)j_l(kr)j_l(kr^\prime)
\phi(r^\prime)Y_{lm}(\hat{\boldsymbol{r}}).\nonumber\\\label{e31}
\end{eqnarray}
If we take $\Omega(p)=1$, then $\Omega(k)=1$, the above equation is simplified as:
\begin{equation}
\phi(r)=\frac{2}{\pi}\int dr^{\prime}{r^{\prime}}^2\int dk\,k^2j_l(kr)j_l(kr^\prime)\phi(r^\prime).\label{e32}
\end{equation}
The above equation is true for arbitrary function $\phi(r)$, which gives the orthogonality condition for the spherical Bessel function
\begin{equation}
\frac{2}{\pi}\,{r^{\prime}}^2\!\int dk\,k^2j_l(kr)j_l(kr^\prime)=\delta(r-r^\prime).\label{e33}
\end{equation}
Furthermore, if we take $\phi(r)=j_l(k^\prime r)$, we can transform  Eq. (\ref{e31}) to a simple form  by using Eq. (\ref{e33}),
\begin{eqnarray}
&&\Omega(p)j_l(k^\prime r)Y_{lm}(\hat{\boldsymbol{r}})\nonumber\\
&&=\int dk\, \Omega(k)j_l(kr)Y_{lm}(\hat{\boldsymbol{r}})\delta(k-k^\prime).\nonumber\\
&&=\Omega(k^\prime)j_l(k^\prime r)Y_{lm}(\hat{\boldsymbol{r}})
\end{eqnarray}
that is
\begin{equation}
\Omega(p)j_l(k r)Y_{lm}(\hat{\boldsymbol{r}})
=\Omega(k)j_l(k r)Y_{lm}(\hat{\boldsymbol{r}}).\label{e34}
\end{equation}
The above equation is actually an eigenequation of $\Omega(p)$. If we expand the unsolved wave function in terms of the spherical Bessel function, the above relation can do a great help in solving the wave equation.

Now we turn to solving  the eigenequation of $H_0$. The spinor part of Eq. (\ref{e26}) can be rewritten as
\begin{equation}
\Psi(\boldsymbol{r})=\left( \begin{array}{cc} \Psi_A\\\Psi_B\end{array}\right )
=\left( \begin{array}{cc} g(r)\;y_{j,l_A}^{m_j}\\i f(r)\;y_{j,l_B}^{m_j}\end{array}\right ), \label{e35}
\end{equation}
by using the formula
\begin{equation}
\boldsymbol{\sigma}\cdot\boldsymbol{p}=i(\boldsymbol{\sigma}\cdot\hat{\boldsymbol{r}})
\left(\frac{\boldsymbol{\sigma}\cdot\boldsymbol{L}}{r}-\frac{d}{dr}\right),
\end{equation}
we have
\begin{eqnarray}
(\boldsymbol{\sigma}\cdot\boldsymbol{p})g(r)\;y_{j,l_A}^{m_j}&=&i\left(\frac{k+1}{r}+\frac{d}{dr}\right)
g(r)\;y_{j,l_B}^{m_j},\label{e42}\\
(\boldsymbol{\sigma}\cdot\boldsymbol{p})f(r)\;y_{j,l_B}^{m_j}&=&-i\left(\frac{k-1}{r}-\frac{d}{dr}\right)
g(r)\;y_{j,l_A}^{m_j},\label{e43}
\end{eqnarray}
where the relations
\begin{eqnarray}
(\boldsymbol{\sigma}\cdot\hat{\boldsymbol{r}})\;y_{j,l_A}^{m_j}=-y_{j,l_B}^{m_j}\\
(\boldsymbol{\sigma}\cdot\hat{\boldsymbol{r}})\;y_{j,l_B}^{m_j}=-y_{j,l_A}^{m_j}
\end{eqnarray}
are used.

Inserting Eq. (\ref{a2}) and (\ref{e37}) into Eq. (\ref{e217}), we can rewrite $H_0$ in the matrix form
\begin{eqnarray}
H_0=\omega_1+\omega_2+\frac{1}{4}\left(1+\frac{H_2}{\omega_2}\right)
(V_v+\beta^{(2)}V_s)\left(1+\frac{H_2}{\omega_2}\right),
\nonumber\\
\end{eqnarray}
where
\begin{eqnarray}
1+\frac{H_2}{\omega_2}
=\left( \begin{array}{cc} \frac{m_2}{\omega_2}+1&\frac{\boldsymbol{\sigma}\cdot\boldsymbol{p}}{\omega_2}  \\ \frac{\boldsymbol{\sigma}\cdot\boldsymbol{p}}{\omega_2}  & -\frac{m_2}{\omega_2}+1 \end{array}\right ),
\\
V_v+\beta^{(2)}V_s=\left( \begin{array}{cc} V_v+V_s& \\  & V_v-V_s \end{array}\right),
\end{eqnarray}
then $H_0$ can be written as:
\begin{eqnarray}
H_0&=&\left( \begin{array}{cc} \omega_1+\omega_2& \\  & \omega_1+\omega_2 \end{array}\right )\nonumber\\
&+&\frac{1}{4}\left( \begin{array}{cc} H_a &H_b \\ H_c & H_d \end{array}\right),\label{e38}
\end{eqnarray}
where
\begin{eqnarray}
H_a&=&(\frac{m_2}{\omega_2}+1)(V_v+V_s)(\frac{m_2}{\omega_2}+1)\nonumber\\
&+&\frac{\boldsymbol{\sigma}\cdot\boldsymbol{p}}{\omega_2}(V_v-V_s)
\frac{\boldsymbol{\sigma}\cdot\boldsymbol{p}}{\omega_2},\\
H_b&=&(\frac{m_2}{\omega_2}+1)(V_v+V_s)\frac{\boldsymbol{\sigma}\cdot\boldsymbol{p}}{\omega_2}\nonumber\\
&+&\frac{\boldsymbol{\sigma}\cdot\boldsymbol{p}}{\omega_2}(V_v-V_s)
(-\frac{m_2}{\omega_2}+1),\\
H_c&=&H_b^{\dagger},\\
H_d&=&(-\frac{m_2}{\omega_2}+1)(V_v-V_s)(-\frac{m_2}{\omega_2}+1)\nonumber\\
&+&\frac{\boldsymbol{\sigma}\cdot\boldsymbol{p}}{\omega_2}(V_v+V_s)
\frac{\boldsymbol{\sigma}\cdot\boldsymbol{p}}{\omega_2}.
\end{eqnarray}

Since the quark and antiquark are bound in the meson, when the distance between them is large enough, the wave
function drops dramatically. The wave function will effectively vanish at a typically large distance $L$, the quark and antiquark can be viewed as being restricted in a limited space, $0 < r < L$. The unsolved functions $f(r)$ and $g(r)$ can be expanded in terms of the spherical Bessel function in the limited space, we write the basis as
\begin{eqnarray}
\psi_i^A&=&\frac{1}{N_i^A}j_{l_A}(\frac{a_i^Ar}{L})\left( \begin{array}{cc} y_{j,l_A}^{m_j} \\0 \end{array}\right ),\label{e39}\\
\psi_\alpha^B&=&\frac{i}{N_\alpha^B}j_{l_B}(\frac{a_\alpha^Br}{L})\left( \begin{array}{cc} 0\\y_{j,l_B}^{m_j}  \end{array}\right ),\label{e40}
\end{eqnarray}
where  the superscripts ``$A$" and ``$B$" stand for the upper and lower parts of Eq. (\ref{e35}), and $N_n$ is the module of the spherical Bessel function
\begin{equation}
N_n^2=\int_0^L dr^\prime r^{\prime 2} j_l(\frac{a_n r^\prime}{L})^2,
\end{equation}
with $a_n$ the $n$-th root of the spherical Bessel function $j_l(x)=0$. $N_n$ is added in accordance with the normalization condition in Eq. (\ref{e44}). In the limited space, the momentum $k$ is discrete, we can have the relevance
\begin{equation}
\frac{a_n }{L} \Longleftrightarrow k.
\end{equation}

The eigenfunction of $H_0$ can be expanded in the orthonormalized basis $\{\psi_i^A, \psi_\alpha^B\}$:
\begin{equation}
\Psi=\sum_{i=1}^\infty g_i\psi_i^A+\sum_{\alpha=1}^\infty f_\alpha \psi_\alpha^B.
\end{equation}
In the numerical calculation the above summation can be truncated at a large integer $N$,
\begin{equation}
\Psi=\sum_{i=1}^N g_i\psi_i^A+\sum_{\alpha=1}^N f_\alpha \psi_\alpha^B.\label{e50}
\end{equation}

According to Eq. (\ref{e38}), we can rewrite the eigenequation of $H_0$ in the representation of $\{\psi_i^A, \psi_\alpha^B\}$. In this representation, the operator $H_0$ has its matrix form
\begin{eqnarray}
H_0&=&\left( \begin{array}{cc} <\omega_1+\omega_2>_{ij}& \\  & <\omega_1+\omega_2>_{\alpha\beta} \end{array}\right )\nonumber\\
&+&\frac{1}{4}\left( \begin{array}{cc} <H_a>_{ij}&<H_b>_{i\beta}\\ <H_c>_{\alpha j} & <H_d>_{\alpha\beta} \end{array}\right ),\label{e41}
\end{eqnarray}

Applying Eqs. (\ref{e34}), (\ref{e42}) and (\ref{e43}) with the normalization condition Eq. (\ref{e45}), we can get the matrix elements of $H_0$ easily.
\begin{eqnarray}
&&<\omega_1+\omega_2>_{ij}=\left[\omega_1(\frac{a_i^A}{L})+\omega_2(\frac{a_i^A}{L})\right]\delta_{ij},\\
&&<\omega_1+\omega_2>_{\alpha\beta}=\left[\omega_1(\frac{a_\alpha^B}{L})+\omega_2(\frac{a_\alpha^B}{L})\right]
\delta_{\alpha\beta},
\end{eqnarray}
here we define a symbolic notation
\begin{equation}
\left<\phi(r)\right>_{m,l_A;n,l_B}=\int_0^L dr\,r^2 j_{l_A}(\frac{a_m^Ar}{L})\phi(r) j_{l_B}(\frac{a_n^Br}{L}),
\end{equation}
then we have
\begin{eqnarray}
&&<H_a>_{ij}=\frac{1}{N_i^AN_j^A}\left(\frac{m_2}{\omega_2(\frac{a_i^A}{L})}+1\right)
\left(\frac{m_2}{\omega_2(\frac{a_j^A}{L})}+1\right)\nonumber\\
&&\times<V_v+V_s>_{i,l_A;j,l_A}\nonumber\\
&&+\frac{1}{N_i^AN_j^A}\frac{1}{\omega_2(\frac{a_i^A}{L})}\frac{1}{\omega_2(\frac{a_j^A}{L})}
\nonumber\\
&&\times\left<\left(\frac{k+1}{r}+\frac{d}{dr}\right)^{\dagger}(V_v-V_s)
\left(\frac{k+1}{r}+\frac{d}{dr}\right)\right>_{i,l_A;j,l_A},\nonumber\\
\end{eqnarray}
\begin{eqnarray}
&&<H_b>_{i\beta}=\frac{1}{N_i^AN_\beta^B}\left(\frac{m_2}{\omega_2(\frac{a_i^A}{L})}+1\right)
\frac{1}{\omega_2(\frac{a_\beta^B}{L})}\nonumber\\
&&\times\left<(V_v+V_s)
\left(\frac{k-1}{r}-\frac{d}{dr}\right)\right>_{i,l_A;\beta,l_B}\nonumber\\
&&+\frac{1}{N_i^AN_\beta^B}\frac{1}{\omega_2(\frac{a_i^A}{L})}
\left(\frac{-m_2}{\omega_2(\frac{a_\beta^B}{L})}+1\right)
\nonumber\\
&&\times\left<\left(\frac{k+1}{r}+\frac{d}{dr}\right)^{\dagger}(V_v-V_s)
\right>_{i,l_A;\beta,l_B},\\
&&<H_c>_{\alpha j}=<H_b>_{j\alpha }^*,
\end{eqnarray}
\begin{eqnarray}
&&<H_d>_{\alpha\beta}=\frac{1}{N_\alpha^BN_\beta^B}\left(\frac{-m_2}{\omega_2(\frac{a_\alpha^B}{L})}+1\right)
\left(\frac{-m_2}{\omega_2(\frac{a_\beta^B}{L})}+1\right)\nonumber\\
&&\times<V_v-V_s>_{\alpha,l_B;\beta,l_B}\nonumber\\
&&+\frac{1}{N_\alpha^BN_\beta^B}\frac{1}{\omega_2(\frac{a_\alpha^B}{L})}\frac{1}{\omega_2(\frac{a_\beta^B}{L})}
\nonumber\\
&&\times\left<\left(\frac{k-1}{r}-\frac{d}{dr}\right)^{\dagger}(V_v+V_s)
\left(\frac{k-1}{r}-\frac{d}{dr}\right)\right>_{\alpha,l_B;\beta,l_B}.\nonumber\\
\end{eqnarray}

Diagonalizing the Hermitian matrix of $H_0$, we can get the eigenenergy of $H_0$ and the coefficients $g_i,f_\alpha$, which are defined in Eq. (\ref{e50}), then the eigenequation associated with $H_0$ is solved and the eigenfunction is obtained.

Now we discuss the perturbative corrections of $H^\prime$.
It is easy to verify that the operators
\begin{equation}
\{{\boldsymbol{j}}^2,j_z,K, S_z \}\nonumber
\end{equation}
are a set of mutually commuting operators which commute with $H_0$, where ${\boldsymbol{j}}=\boldsymbol{L}+\boldsymbol{S}^{(2)}$, $\boldsymbol{S}^{(2)}=\frac{1}{2}
\boldsymbol{\Sigma}^{(2)}$, $K=\beta^{(2)}(\boldsymbol{\Sigma}^{(2)}\cdot\boldsymbol{L}+1)$, $\boldsymbol{S}=\frac{1}{2}\boldsymbol{\sigma}^{(1)}$. Then the eigenstates of $H_0$ can be labeled by the corresponding set of quantum numbers $\{n,j,m_j,k,s\}$ and the eigenequation associated with $H_0$ can be written as
\begin{equation}
H_0\Psi^{(0)}_{n,k,j,m_j,s}(\boldsymbol{r})=E^{(0)}_{n,k,j}\Psi^{(0)}_{n,k,j,m_j,s}(\boldsymbol{r}),
\end{equation}
where
\begin{equation}
\Psi^{(0)}_{n,k,j,m_j,s}(\boldsymbol{r})=\left( \begin{array}{cc} g_{n,l,j}(r)y_{j,l}^{m_j}(\theta,\varphi)\\i f_{n,l,j}(r)y_{j,2j-l}^{m_j}(\theta,\varphi)\end{array}\right )\chi_s \label{e26}
\end{equation}
with
\begin{equation}
y_{j,l}^m=\left( \begin{array}{cc}k_{j,l,m}^+Y_l^{m-1/2}
\\k_{j,l,m}^-Y_l^{m+1/2}\end{array}\right ),\label{a1}
\end{equation}
\begin{equation}
k_{j,l,m}^\pm=\left\{\begin{array}{ll}
+\sqrt{\frac{l\pm m+1/2}{2l+1}},\; j=l+1/2 \\
\mp\sqrt{\frac{l\mp m+1/2}{2l+1}},\; j=l-1/2.
 \end{array}\right.
\end{equation}
In Eq. (\ref{a1}), $Y_l^{m\pm1/2}$ is the spherical harmonics. For $j=l\pm1/2$, we have
\begin{equation}
\int d\Omega\,(y_{j,l}^{m_j})^\dagger y_{j,l}^{m_j}=1.\label{e45}
\end{equation}

Thus the normalization condition of Eq. (\ref{e26}) is read as
\begin{equation}
\int_0^\infty\!dr\,r^2 \,(f_{n,l,j}^2+g_{n,l,j}^2)=1.\label{e44}\\
\end{equation}

For a state with quantum number $j$, the operator $K$ can have two opposite eigenvalues $\pm(j+1/2)$ with
\begin{equation}
l=\left\{\begin{array}{ll}
j+1/2, \; k=+(j+1/2) \\
j-1/2, \;k=-(j+1/2),
 \end{array}\right.
\end{equation}
where the quantum number $k$ is defined as
\begin{equation}
K\Psi^{(0)}_{n,k,j,m_j,s}(\boldsymbol{r})=-k\Psi^{(0)}_{n,k,j,m_j,s}(\boldsymbol{r}).
\end{equation}
The zeroth order invariant mass $E^0_{n,k,j}$ is equivalently determined by $n,j,l$, and the eigenstate has the parity $P=(-1)^{l+1}$. The Solution of the eigenequation associated with $H_0$ is detailed in the next section.

The perturbative term $H^\prime$ does not commute with any of the operators introduced in solving the eigenequation of $H_0$, but still commutes with
\begin{equation}
\{{\boldsymbol{J}}^2,J_z,\mathcal{P}\}\nonumber
\end{equation}
where $\boldsymbol{J}=\boldsymbol{j}+\boldsymbol{S}$ and $\mathcal{P}$ is the parity operator. Thus the eigenstates of the total Hamiltonian $H=H_0+H^\prime$ can be labeled by the set of quantum numbers $\{n,J,M_J,P\}$.

With the help of Clebsch-Gordan coefficients, we can compose the basis states with the quantum numbers set  $\{n,k,j,J,M_J\}$ by combining the eigenstates of $H_0$,
\begin{equation}
\Psi^{(0)}_{n,k,j;J,M_J}(\boldsymbol{r})=\sum_{m_j,s} C^{J,M_J}_{j,m_j;1/2,s}
\Psi^{(0)}_{n,k,j;m_j,s}(\boldsymbol{r}),\label{ee1}
\end{equation}
then calculate the corrections and mixings in the obtained basis.

The correction in first order perturbation can be written as
\begin{equation}
E_{n,l,j,J}=E^{(0)}_{n,l,j}+\frac{1}{m_1}\delta E^{(1)}_{n,l,j,J}
\end{equation}

Mixing can happen between states with same quantum numbers $J,M_J,P$, but with different quantum number $j$. In this paper the mixing between states with $j=l\pm1/2$ is considered. Mixing can be described by the mass matrix. The mass matrix is calculated perturbatively in the basis, which can be written as
\begin{equation}
H=\left( \begin{array}{cc} H_{11}& H_{12}\\ H_{21} & H_{22} \end{array}\right ),
\end{equation}
where $H_{ij}=\langle \Psi_i^{(0)} |H|\Psi_j^{(0)}\rangle$, with $i, j=1, 2$, $\Psi_{i,j}^{(0)}$'s denote the basis states which are defined in Eq. (\ref{ee1}).

In Eq.(\ref{e218}), there is a term $(1+h_2)/2$ on both left and right sides of the expression. We can let $(1+h_2)/2$ operate on the wave functions of $H_0$ and obtain new wave functions, then deal with the middle term of Eq.(\ref{e218}) with the new wave functions.
\begin{eqnarray}
&&\frac{1}{2}(1+h_2)\left( \begin{array}{cc} g(r)\;y_{j,l_A}^{m_j}\\i f(r)\;y_{j,l_B}^{m_j}\end{array}\right )\nonumber\\
&&=\frac{1}{2}\left( \begin{array}{cc} \frac{m_2}{\omega_2}+1&\frac{\boldsymbol{\sigma}\cdot\boldsymbol{p}}{\omega_2} \\ \frac{\boldsymbol{\sigma}\cdot\boldsymbol{p}}{\omega_2} & -\frac{m_2}{\omega_2}+1\end{array}\right )\left( \begin{array}{cc} g(r)\;y_{j,l_A}^{m_j}\\i f(r)\;y_{j,l_B}^{m_j}\end{array}\right )\nonumber\\
&&=\left( \begin{array}{cc} [\frac{1}{2}(\frac{m_2}{\omega_2}+1)g(r)+\frac{1}{2\omega_2}(\frac{\kappa-1}{r}-\frac{d}{dr})f(r)]\;y_{j,l_A}^{m_j}
\\i [\frac{1}{2\omega_2}(\frac{\kappa+1}{r}+\frac{d}{dr})g(r)+\frac{1}{2}(-\frac{m_2}{\omega_2}+1)f(r)]\;y_{j,l_B}^{m_j}\end{array}\right )\nonumber\\
&&\Rightarrow\left( \begin{array}{cc} g(r)\;y_{j,l_A}^{m_j}\\i f(r)\;y_{j,l_B}^{m_j}\end{array}\right ),
\end{eqnarray}
the above equation shows the transformation from the wave functions of $H_0$ to the
new wave functions with the operation of $(1+h_2)/2$.

With the new wave functions the perturbative contributions of the $H^\prime$ to the eigenvalues of states can be calculated easily, they are given below:

(1) $J^P=0^-$

Only state with $j=1/2$, $l=0$ can construct the $J^P=0^-$ state, we have
\begin{eqnarray}
&&E_{n,0,\frac{1}{2},0}=E_{n,0,\frac{1}{2}}\nonumber\\
&&-\frac{2}{m_1}\int_0^{\infty}\! V_v \left(2f_{n,0,\frac{1}{2}}+r f_{n,0,\frac{1}{2}}^\prime\right)g_{n,0,\frac{1}{2}}\;r\,dr \nonumber\\
\end{eqnarray}

(2) $J^P=1^-$

Both states with $j=1/2, l=0$ and $j=3/2, l=2$ can construct the $J^P=1^-$ state, we have
\begin{eqnarray}
&&E_{n,0,\frac{1}{2},1}=E_{n,0,\frac{1}{2}}\nonumber\\
&&-\frac{2}{3m_1}\int_0^{\infty}\! V_v \left(g_{n,0,\frac{1}{2}}f_{n,0,\frac{1}{2}}^\prime-2f_{n,0,\frac{1}{2}}g_{n,0,\frac{1}{2}}^\prime
\right)\;r^2\,dr \nonumber\\
\end{eqnarray}
and
\begin{eqnarray}
&&E_{n,2,\frac{3}{2},1}=E_{n,2,\frac{3}{2}}+\frac{1}{3m_1}\int_0^{\infty}\! V_v\left(12\,g_{n,2,\frac{3}{2}}f_{n,2,\frac{3}{2}} \right.\nonumber\\
&&\left.-r g_{n,2,\frac{3}{2}}f_{n,2,\frac{3}{2}}^\prime+5\,r f_{n,2,\frac{3}{2}}g_{n,2,\frac{3}{2}}^\prime
\right)\;r\,dr
\end{eqnarray}

(3) $J^P=0^+$

Only state with $j=1/2$, $l=1$ can construct the $J^P=0^+$ state, we have
\begin{eqnarray}
&&E_{n,1,\frac{1}{2},0}=E_{n,1,\frac{1}{2}}\nonumber\\
&&+\frac{2}{m_1}\int_0^{\infty}\! V_v \left(2g_{n,1,\frac{1}{2}}+r g_{n,1,\frac{1}{2}}^\prime\right)f_{n,1,\frac{1}{2}}\;r\,dr \nonumber\\
\end{eqnarray}

(4) $J^P=1^+$

Both states with $j=1/2, l=1$ and $j=3/2, l=1$ can construct the $J^P=1^+$ state, the matrix elements of the mass
matrix are
\begin{eqnarray}
&&H_{11}=E_{n,1,\frac{1}{2}}\nonumber\\
&&+\frac{2}{3m_1}\int_0^{\infty}\! V_v \left(f_{n,1,\frac{1}{2}}g_{n,1,\frac{1}{2}}^\prime-2\,g_{n,1,\frac{1}{2}}f_{n,1,\frac{1}{2}}^\prime
\right)\;r^2\,dr, \nonumber\\
\\
&&H_{12}=\frac{1}{3\sqrt{2}m_1}\int_0^{\infty}\! V_v\left(3\,f_{n,1,\frac{1}{2}}g_{n,1,\frac{3}{2}}
 \right.\nonumber\\&&\left.
 +3\,f_{n,1,\frac{3}{2}}g_{n,1,\frac{1}{2}} +r f_{n,1,\frac{1}{2}}g_{n,1,\frac{3}{2}}^\prime+r f_{n,1,\frac{3}{2}}g_{n,1,\frac{1}{2}}^\prime
 \right.\nonumber\\&&\left.
 +r g_{n,1,\frac{1}{2}}f_{n,1,\frac{3}{2}}^\prime+r g_{n,1,\frac{3}{2}}f_{n,1,\frac{1}{2}}^\prime
\right)\;r\,dr,  \\
&&H_{21}=H_{12}^*,\\
&&H_{22}=E_{n,1,\frac{3}{2}}-\frac{1}{3m_1}\int_0^{\infty}\! V_v\left(12\,f_{n,1,\frac{3}{2}}g_{n,1,\frac{3}{2}} \right.\nonumber\\
&&\left.-r f_{n,1,\frac{3}{2}}g_{n,1,\frac{3}{2}}^\prime+5\,r g_{n,1,\frac{3}{2}}f_{n,1,\frac{3}{2}}^\prime
\right)\;r\,dr.
\end{eqnarray}
With the matrix elements given above, one can get the eigenvalues of the two mixing states and
the mixing angle easily by  diagonalizing the mass matrix.

(5) $J^P=2^+$

Both states with $j=3/2, l=1$ and $j=5/2, l=3$ can construct the $J^P=2^+$ state, we have
\begin{eqnarray}
&&E_{n,1,\frac{3}{2},2}=E_{n,1,\frac{3}{2}}\nonumber\\
&&-\frac{1}{5m_1}\int_0^{\infty}\! V_v \left(4\,f_{n,1,\frac{3}{2}}g_{n,1,\frac{3}{2}}
-7\,r f_{n,1,\frac{3}{2}}g_{n,1,\frac{3}{2}}^\prime
 \right.\nonumber\\&&\left.
+3\,r g_{n,1,\frac{3}{2}}f_{n,1,\frac{3}{2}}^\prime
\right)\;r\,dr
\end{eqnarray}
and
\begin{eqnarray}
&&E_{n,3,\frac{5}{2},2}=E_{n,3,\frac{5}{2}}\nonumber\\
&&+\frac{2}{5m_1}\int_0^{\infty}\! V_v \left(12\,f_{n,3,\frac{5}{2}}g_{n,3,\frac{5}{2}}
+4\,r f_{n,3,\frac{5}{2}}g_{n,3,\frac{5}{2}}^\prime
 \right.\nonumber\\&&\left.
-\,r g_{n,3,\frac{5}{2}}f_{n,3,\frac{5}{2}}^\prime
\right)\;r\,dr
\end{eqnarray}

(6) $J^P=2^-$

Both states with $j=3/2, l=2$ and $j=5/2, l=2$ can construct the $J^P=2^-$ state, the matrix elements of the mass
matrix are
\begin{eqnarray}
&&H_{11}=E_{n,2,\frac{3}{2}}\nonumber\\
&&+\frac{1}{5m_1}\int_0^{\infty}\! V_v \left(4\,f_{n,2,\frac{3}{2}}g_{n,2,\frac{3}{2}}
-7\,r g_{n,2,\frac{3}{2}}f_{n,2,\frac{3}{2}}^\prime
 \right.\nonumber\\&&\left.
+3\,r f_{n,2,\frac{3}{2}}g_{n,2,\frac{3}{2}}^\prime
\right)\;r\,dr, \\
&&H_{12}=\frac{1}{5m_1}\sqrt{\frac{3}{2}}\int_0^{\infty}\! V_v\left(3\,f_{n,2,\frac{3}{2}}g_{n,2,\frac{5}{2}}
 \right.\nonumber\\&&\left.
 +3\,f_{n,2,\frac{5}{2}}g_{n,2,\frac{3}{2}} +r f_{n,2,\frac{3}{2}}g_{n,2,\frac{5}{2}}^\prime+r f_{n,2,\frac{5}{2}}g_{n,2,\frac{3}{2}}^\prime
 \right.\nonumber\\&&\left.
 +r g_{n,2,\frac{3}{2}}f_{n,2,\frac{5}{2}}^\prime+r g_{n,2,\frac{5}{2}}f_{n,2,\frac{3}{2}}^\prime
\right)\;r\,dr,  \\
&&H_{21}=H_{12}^*,\\
&&H_{22}=E_{n,2,\frac{5}{2}}-\frac{2}{5m_1}\int_0^{\infty}\! V_v \left(12\,f_{n,2,\frac{5}{2}}g_{n,2,\frac{5}{2}}
 \right.\nonumber\\&&\left.
 +4\,r g_{n,2,\frac{5}{2}}f_{n,2,\frac{5}{2}}^\prime
-\,r f_{n,2,\frac{5}{2}}g_{n,2,\frac{5}{2}}^\prime
\right)\;r\,dr.
\end{eqnarray}
By diagonalizing the mass matrix, one can get the eigenvalues of the two mixing states and
the mixing angle.

(7) $J^P=3^-$

Both states with $j=5/2, l=2$ and $j=7/2, l=4$ can construct the $J^P=3^-$ state, we have
\begin{eqnarray}
&&E_{n,2,\frac{5}{2},2}=E_{n,2,\frac{5}{2}}\nonumber\\
&&-\frac{2}{7m_1}\int_0^{\infty}\! V_v \left(6\,f_{n,2,\frac{5}{2}}g_{n,5,\frac{5}{2}}
-5\,r f_{n,2,\frac{5}{2}}g_{n,2,\frac{5}{2}}^\prime
 \right.\nonumber\\&&\left.
+2\,r g_{n,2,\frac{5}{2}}f_{n,2,\frac{5}{2}}^\prime
\right)\;r\,dr
\end{eqnarray}
and
\begin{eqnarray}
&&E_{n,4,\frac{7}{2},2}=E_{n,4,\frac{7}{2}}\nonumber\\
&&+\frac{1}{7m_1}\int_0^{\infty}\! V_v \left(40\,f_{n,4,\frac{7}{2}}g_{n,4,\frac{7}{2}}
+11\,r f_{n,4,\frac{7}{2}}g_{n,4,\frac{7}{2}}^\prime
 \right.\nonumber\\&&\left.
-3\,r g_{n,4,\frac{7}{2}}f_{n,4,\frac{7}{2}}^\prime
\right)\;r\,dr
\end{eqnarray}

(8) $J^P=3^+$

Both states with $j=5/2, l=3$ and $j=7/2, l=3$ can construct the $J^P=3^+$ state, the matrix elements of the mass
matrix are
\begin{eqnarray}
&&H_{11}=E_{n,3,\frac{5}{2}}\nonumber\\
&&+\frac{2}{7m_1}\int_0^{\infty}\! V_v \left(6\,f_{n,3,\frac{5}{2}}g_{n,3,\frac{5}{2}}
-5\,r g_{n,3,\frac{5}{2}}f_{n,3,\frac{5}{2}}^\prime
 \right.\nonumber\\&&\left.
+2\,r f_{n,3,\frac{5}{2}}g_{n,3,\frac{5}{2}}^\prime
\right)\;r\,dr, \\
&&H_{12}=\frac{\sqrt{3}}{7m_1}\int_0^{\infty}\! V_v\left(3\,f_{n,3,\frac{5}{2}}g_{n,3,\frac{7}{2}}
 \right.\nonumber\\&&\left.
 +3\,f_{n,3,\frac{7}{2}}g_{n,3,\frac{5}{2}} +r f_{n,3,\frac{5}{2}}g_{n,3,\frac{7}{2}}^\prime+r f_{n,3,\frac{7}{2}}g_{n,3,\frac{5}{2}}^\prime
 \right.\nonumber\\&&\left.
 +r g_{n,3,\frac{5}{2}}f_{n,3,\frac{7}{2}}^\prime+r g_{n,3,\frac{7}{2}}f_{n,3,\frac{5}{2}}^\prime
\right)\;r\,dr,  \\
&&H_{21}=H_{12}^*,\\
&&H_{22}=E_{n,3,\frac{7}{2}}-\frac{1}{7m_1}\int_0^{\infty}\! V_v \left(40\,f_{n,3,\frac{7}{2}}g_{n,3,\frac{7}{2}}
 \right.\nonumber\\&&\left.
 +11\,r g_{n,3,\frac{7}{2}}f_{n,3,\frac{7}{2}}^\prime
-3\,r f_{n,3,\frac{7}{2}}g_{n,3,\frac{7}{2}}^\prime
\right)\;r\,dr.
\end{eqnarray}
The eigenvalues of the two mixing states and the mixing angle can be obtained by diagonalizing the mass matrix.

Once the  eigenfuncion of $H_0$ is obtained, the perturbative correction of $H^\prime$ can be calculated directly by following the procedures presented above.

\section*{IV Numerical result and discussion}
The parameters used in this work are the quark masses, $m_{u,d}$, $m_s$, $m_c$, $m_b$ and three potential parameters $\alpha_s$, $b$, $c$. In the calculation of the spectra, we find that the model is capable to give reasonable spectral structure for each meson system. In the framework of Bethe-Salpeter equation, the parameter $b$ is responsible for the energy levels of states with higher quantum numbers $n$ or $l$. We find that if the parameter $b$ is taken as a constant for all the meson systems, the energy levels for the excited meson states decrease as the value of $m_2$ decreases. As a result, if we determine the value of $b$ for $D_s$, $B_s$ mesons, then calculate $D$, $B$ mesons with the same value of $b$ but smaller value of $m_2$, we can not get the correct energy levels for the excited states of $D$, $B$ mesons. This can be explained by the form of $H_0$. Unlike the Dirac-like Hamiltonian, the influence of the confining potential $V_s(r)$ weakens as $m_2$ decreases in Eq. (\ref{e217}). This difficulty suggests that the parameter $b$ may depend on the masses of the quark and antiquark, especially the lighter one of them.

With the considerations above, our fitting of the parameters gives the following values
\begin{eqnarray}
&&  m_{u,d}=0.360\; {\rm GeV},\nonumber\\
&&m_s=0.550\;{\rm GeV},\nonumber\\
&&m_c=1.478\; {\rm GeV},\nonumber\\
&& m_b=4.865\; {\rm GeV},\nonumber\\
&&\alpha_s=0.513,\nonumber\\
&&b=\left\{\begin{array}{ll}
 0.350\;{\rm GeV}^2& {\rm for}\; (c\bar{q},b\bar{q})\; {\rm system}, \\
 0.260\;{\rm GeV}^2& {\rm for}\; (c\bar{s},b\bar{s})\; {\rm system},
 \end{array}\right.\nonumber\\
&&c=-0.320\; {\rm GeV}.\nonumber
\end{eqnarray}

Numerical calculation shows that the solution of the wave equation is stable when $L>5$ fm, $N>35$. Here we take $L=10$ fm, $N=50$. We fit the spectra of the heavy-light $D$, $D_s$, $B$, $B_s$ mesons mainly based on the meson states presented in PDG \cite{PDG}. The numerical results for the spectra of heavy-light mesons are presented in two tables. Table I is for $D$, $D_s$ mesons and Table II for $B$, $B_s$ mesons. The obtained spectra are in reasonable agreement with the experimental measurements.
Theoretical deviations from experimental data mainly appear in the $D_s$ meson sector, where the calculated mass of the $D_{s0}^*(2317)^0$ resonance is the worst. In the two tables, one can see that apart from the $D_{s0}^*(2317)^0$ and $D_{s1}(2460)$ resonances, mass calculations for the rest of the resonances are in good agreement with experimental data.
 The discrepancy may be ascribed to the naive assumption of the kernel, the dropped $\alpha^2_s$ term in the Hamiltonian can also make the result incomplete. The highly excited meson states are also calculated in the spectra and the newly observed charmed meson states are identified in our model.

\begin{table*}[!htbp]
\caption{Spectrum for $D$ and $D_s$ mesons. $E^0$ denotes the lowest order energies. $E^\mathrm{phys.}$ includes all the corrections of order $1/m_Q$. All units are in MeV.}
 \label{t1}
\begin{tabular}{c||c|c|c|c||c|c|c|c}\hline
$n^jL_J    $ & Meson         &$E_\mathrm{expt.}$ \cite{PDG,LHCb}& $E^0$ & $E^\mathrm{phys.}$&Meson&$E_\mathrm{expt.}$ \cite{PDG}& $E^0$ & $E^\mathrm{phys.}$\\ \hline
$1^{1/2}S_0$ &  $D$          &$1869.62\pm0.15$  & 2105  & 1859&$D_s^\pm$      &$1968.49\pm 0.32$ & 2179  & 1949  \\
$1^{1/2}S_1$ &  $D^*$        &$2010.28\pm0.13$  & 2105  & 2026&$D_s^{*\pm}$   & $2112.3\pm 0.5$  & 2179  & 2110  \\

$1^{1/2}P_0$ &$D_0^*(2400)^0$&$2318 \pm 29   $  & 2644  & 2357&$D_{s0}^*(2317)^0$&$2317.8\pm 0.6$ & 2659  & 2412  \\
$1^{1/2}P_1$ &               &                  & 2644  & 2529&$D_{s1}(2536)$ & $2535.12\pm 0.13$ & 2659  & 2562  \\
$1^{3/2}P_1$ &$D_1(2420)$    &$2421.3\pm0.6  $  & 2530  & 2434&$D_{s1}(2460)$ &$2459.6\pm 0.6$   & 2618  & 2528  \\
$1^{3/2}P_2$ &$D_2^*(2460)$  &$2464.4 \pm1.9 $  & 2530  & 2482&$D_{s2}^*(2573)$& $2571.9\pm 0.8$ & 2618  & 2575  \\

$1^{3/2}D_1$ &               &                  & 2968  & 2852&$D_{s1}^*(2860)^-$&$2859\pm12\pm6\pm23$ \cite{LHCbspin13b}   & 2976  & 2873  \\
$1^{3/2}D_2$ &               &                  & 2968  & 2900&               &                  & 2976  & 2916  \\
$1^{5/2}D_2$ &$D_J(2740)^0$  &$2737.0\pm 3.5\pm11.2$& 2793  & 2728&                &                  & 2889  & 2829  \\
$1^{5/2}D_3$ &$D_J^*(2760)^0$&$2760.1\pm 1.1\pm3.7$& 2793   & 2753&$D_{s3}^*(2860)^-$&$2860.5\pm2.6\pm2.5\pm6.0$ \cite{LHCbspin13b}& 2889  & 2852  \\

$1^{5/2}F_2$ &               &                  & 3188  & 3107&               &                  & 3201  & 3128  \\
$1^{5/2}F_3$ &               &                  & 3188  & 3134&               &                  & 3201  & 3152  \\
$1^{7/2}F_3$ &$D_J(3000)^0$  &$2971.8\pm 8.7$   & 2996  & 2942&               &                  & 3098  & 3049  \\

$2^{1/2}S_0$ &$D_J(2580)^0$  &$2579.5\pm 3.4\pm 5.5$& 2821  & 2575&$D_{sJ}(2632)$& $2632.5\pm 1.7$ \cite{Ds2632}           & 2849  & 2624  \\
$2^{1/2}S_1$ &$D_J^*(2650)^0$&$2649.2\pm 3.5\pm 3.5$& 2821  & 2686&$D_{s1}^*(2710)$ & $2708\pm9^{+11}_{-10}$ \cite{Ds2700}  & 2849  & 2729  \\

$2^{1/2}P_0$ &               &                  & 3133  & 2902&               &                  & 3136  & 2918  \\
$2^{1/2}P_1$ &               &                  & 3133  & 2999&$D_{sJ}(3040)$ &$3044\pm8^{+30}_{-5}$ \cite{Ds2009} & 3136  & 3017 \\
$2^{3/2}P_1$ &               &                  & 3052  & 2932&               &                  & 3106  & 2994  \\
$2^{3/2}P_2$ &$D^*_J(3000)^0$&$3008.1\pm 4.0$   & 3052  & 2969&               &                  & 3106  & 3031  \\

$2^{3/2}D_1$ &               &                  & 3350  & 3228&               &                  & 3360  & 3247 \\
$2^{3/2}D_2$ &               &                  & 3350  & 3260&               &                  & 3360  & 3278  \\
$2^{5/2}D_2$ &               &                  & 3225  & 3139&               &                  & 3296  & 3217  \\
$2^{5/2}D_3$ &               &                  & 3225  & 3160&               &                  & 3296  & 3237  \\

$2^{5/2}F_2$ &               &                  & 3516  & 3425&               &                  & 3534  & 3449  \\
$2^{5/2}F_3$ &               &                  & 3516  & 3444&               &                  & 3534  & 3468  \\
$2^{7/2}F_3$ &               &                  & 3371  & 3301&               &                  & 3455  & 3390  \\
\hline
\end{tabular}
\end{table*}

\begin{table*}[!htbp]
\caption{Spectrum for $B$ and $B_s$ mesons. $E^0$ denotes the lowest order energies. $E^\mathrm{phys.}$ includes all the corrections of order $1/m_Q$. All units are in MeV.}
 \label{t2}
\begin{tabular}{c||c|c|c|c||c|c|c|c}\hline
$n^jL_J    $ & Meson         &$E_\mathrm{expt.}$ \cite{PDG}& $E^0$ & $E^\mathrm{phys.}$&Meson&$E_\mathrm{expt.}$ \cite{PDG}& $E^0$ & $E^\mathrm{phys.}$\\ \hline
$1^{1/2}S_0$ &  $B$           &$5279.25\pm 0.17$& 5362  & 5262&$B_s$          &$5366.77\pm 0.24$      & 5435  & 5337  \\
$1^{1/2}S_1$ &  $B^*$         &$5325.2\pm 0.4$  & 5362  & 5330&$B_s^*$        &$5415.4^{+2.4}_{-2.1} $& 5435  & 5405  \\

$1^{1/2}P_0$ &               &                  & 5859  & 5740&               &                  & 5883  & 5776  \\
$1^{1/2}P_1$ &               &                  & 5859  & 5812&               &                  & 5883  & 5841  \\
$1^{3/2}P_1$ &$B_1(5721)$    &$5723.5\pm 2.0$   & 5772  & 5736&$B_{s1}(5830)$ &$5829.4\pm 0.7$   & 5860  & 5824  \\
$1^{3/2}P_2$ &$B_2^*(5747)$  & $5743\pm 5$      & 5772  & 5754&$B_{s2}^*(5840)$&$5839.7\pm 0.6$  & 5860  & 5843  \\

$1^{3/2}D_1$ &               &                  & 6174  & 6128&               &                  & 6188  & 6146  \\
$1^{3/2}D_2$ &               &                  & 6174  & 6147&               &                  & 6188  & 6163  \\
$1^{5/2}D_2$ &               &                  & 6013  & 5989&               &                  & 6108  & 6085  \\
$1^{5/2}D_3$ &               &                  & 6013  & 5998&               &                  & 6108  & 6094  \\

$1^{5/2}F_2$ &               &                  & 6375  & 6344&               &                  & 6392  & 6363  \\
$1^{5/2}F_3$ &               &                  & 6375  & 6354&               &                  & 6392  & 6373  \\
$1^{7/2}F_3$ &               &                  & 6195  & 6175&               &                  & 6295  & 6276  \\

$2^{1/2}S_0$ &               &                  & 6013  & 5915&               &                  & 6052  & 5961  \\
$2^{1/2}S_1$ &               &                  & 6013  & 5959&               &                  & 6052  & 6003  \\

$2^{1/2}P_0$ &               &                  & 6302  & 6211&               &                  & 6315  & 6227  \\
$2^{1/2}P_1$ &               &                  & 6302  & 6249&               &                  & 6315  & 6266 \\
$2^{3/2}P_1$ &               &                  & 6232  & 6189&               &                  & 6293  & 6249  \\
$2^{3/2}P_2$ &               &                  & 6232  & 6200&               &                  & 6293  & 6263  \\

$2^{3/2}D_1$ &               &                  & 6506  & 6458&               &                  & 6523  & 6478 \\
$2^{3/2}D_2$ &               &                  & 6506  & 6471&               &                  & 6523  & 6491  \\
$2^{5/2}D_2$ &               &                  & 6389  & 6357&               &                  & 6465  & 6434  \\
$2^{5/2}D_3$ &               &                  & 6389  & 6365&               &                  & 6465  & 6441  \\

$2^{5/2}F_2$ &               &                  & 6656  & 6621&               &                  & 6680  & 6647  \\
$2^{5/2}F_3$ &               &                  & 6656  & 6629&               &                  & 6680  & 6654  \\
$2^{7/2}F_3$ &               &                  & 6519  & 6493&               &                  & 6605  & 6580  \\
\hline
\end{tabular}
\end{table*}

As for $D$ mesons, several resonances are observed by LHCb collaboration in the mass region between 2500 and 2800 MeV as well as the region around 3000 Mev, their masses are measured as \cite{LHCb}
\begin{eqnarray}
 M(D_J(2580)^0)&=&2579.5\pm3.4\pm5.5\;\mbox{MeV}\nonumber\\
 M(D_J^*(2650)^0)&=&2649.2\pm3.5\pm3.5\;\mbox{MeV}\nonumber\\
 M(D_J(2740)^0)&=&2737.0\pm3.5\pm11.2\;\mbox{MeV}\nonumber\\
 M(D_J^*(2760)^0)&=&2760.1\pm1.1\pm3.7\;\mbox{MeV}\nonumber\\
 M(D_J^*(2760)^+)&=&2771.7\pm1.7\pm3.8\;\mbox{MeV}\nonumber\\
 M(D_J(3000)^0)&=&2971.8\pm8.7\;\mbox{MeV} \nonumber\\
 M(D_J^*(3000)^0)&=&3008.1\pm4.0\;\mbox{MeV}\nonumber\\
 M(D_J^*(3000)^+)&=&3008.1\;\mbox{MeV}.\nonumber
 \end{eqnarray}
The assignments of the above observed states are listed in Table I. In the calculated spectrum of $D$ meson, $D_J(2740)$ can be identified as the $|1^{5/2}D_2\rangle$ state with $J^P=2^-$ and $D_J^*(2760)$ can be identified as the $|1^{5/2}D_3\rangle$ state with $J^P=3^-$, which agrees with Ref. \cite{CFGN}. In Ref. \cite{LHCb}, $D_J(2580)$ and $D_J^*(2650)$ are identified as $n=2$ excited states with $J^P=0^-$ and $J^P=1^-$, respectively. Our result favors this assignment with $D_J(2580)$ assigned as the $|2^{1/2}S_0\rangle$ state and $D_J^*(2650)$ as the $|2^{1/2}S_1\rangle$ state. One can see that $D_J^*(3000)$ and $D_J(3000)$ can be assigned in Table I as the $|2^{1/2}P_1\rangle$ and $|2^{3/2}P_2\rangle$ states, respectively. But as presented in Ref. \cite{LHCb}, $D_J^*(3000)$ favors the natural parity and $D_J(3000)$ the unnatural parity. With this consideration, $D_J^*(3000)$ is identified as the $|2^{3/2}P_2\rangle$ state with $J^P=2^+$ and $D_J(3000)$ as the $|1^{7/2}F_3\rangle$ state with $J^P=3^+$ in this work.

As for $D_s$ mesons, Several new $D_s$ meson states have been observed besides the low-lying states, such as $D_{sJ}(2632)$ \cite{Ds2632}, $D_{sJ}^*(2860)$ \cite{Ds2860}, $D_{s1}^*(2710)$ \cite{Ds2700}, $D_{sJ}(3040)$ \cite{Ds2009}. $D_{sJ}(2632)$ is measured to be $2632.5\pm1.7$ MeV in Ref. \cite{Ds2632}, we assign it as the $|2^{1/2}S_0\rangle$ state with $J^P=0^-$. For $D_{s1}^*(2710)$, it is measured to be $2708\pm9^{+11}_{-10}$ Mev in Ref. \cite{Ds2700}, it can be assigned as the $|2^{1/2}S_1\rangle$ state with $J^P=1^-$, which agrees with Refs. \cite{CFNR,CTLS}. The $D_{sJ}(3040)$ resonance is observed by BABAR Collaboration at a mass of $3044\pm8_{-5}^{+30}$ Mev \cite{Ds2009}. Several possible assignments of this resonance is discussed in Ref. \cite{CF1}. It is identified as the $|2^{1/2}P_1\rangle$ state with $J^P=1^+$ in our predicted mass spectrum for $D_s$ meson. Recently, LHCb collaboration identifies the resonance $D_{sJ}^*(2860)$ as an admixture of spin-1 and spin-3 resonances and determines the masses and widths of the two states to be \cite{LHCbspin13b,LHCbspin13a}
\begin{eqnarray}
M(D_{s1}^*(2860)^-)&=&2859\pm12\pm6\pm23\;\mbox{MeV},\nonumber\\
\Gamma(D_{s1}^*(2860)^-)&=&159\pm23\pm27\pm72\;\mbox{MeV},\nonumber\\
M(D_{s3}^*(2860)^-)&=&2860.5\pm2.6\pm2.5\pm6.0\;\mbox{MeV},\nonumber\\
\Gamma(D_{s3}^*(2860)^-)&=&53\pm7\pm4\pm6\;\mbox{MeV}.\nonumber
\end{eqnarray}
In Table I, One can see that our predictions for the masses of the $|1^{3/2}D_1\rangle$ and $|1^{5/2}D_3\rangle$ states are around 2860 Mev. Our model appears to be able to interpret the two states being the $J^P=1^-$ and $3^-$ members of the 1D family. We can assign $D_{s1}^*(2860)^-$ as the $|1^{3/2}D_1\rangle$ state with $J^P=1^-$ and $D_{s3}^*(2860)^-$ as the $|1^{5/2}D_3\rangle$ state with $J^P=3^-$.

The wave function of each bound state can be obtained simultaneously when solving the wave equation. The radial wave functions $g_{n,l,j}(r)$ and $f_{n,l,j}(r)$ for $D$ meson are depicted in Fig.1 as an example. $[g_{n,l,j}^2(r)+f_{n,l,j}^2(r)]r^2$ is the possibility density distributed along the quark-antiquark distance $r$. In Ref. \cite{PE}, the radial wave functions are also presented, they are calculated in a model derived by reducing the spectator equation in relation to the Bethe-Salpeter equation \cite{JZ}. One can find that their wave functions share the same pattern with ours, though the  Hamiltonian $H_0$ in our model looks very different from the Dirac-like Hamiltonian used by them.

\begin{figure*}
\centering
\scalebox{1}{\epsfig{file=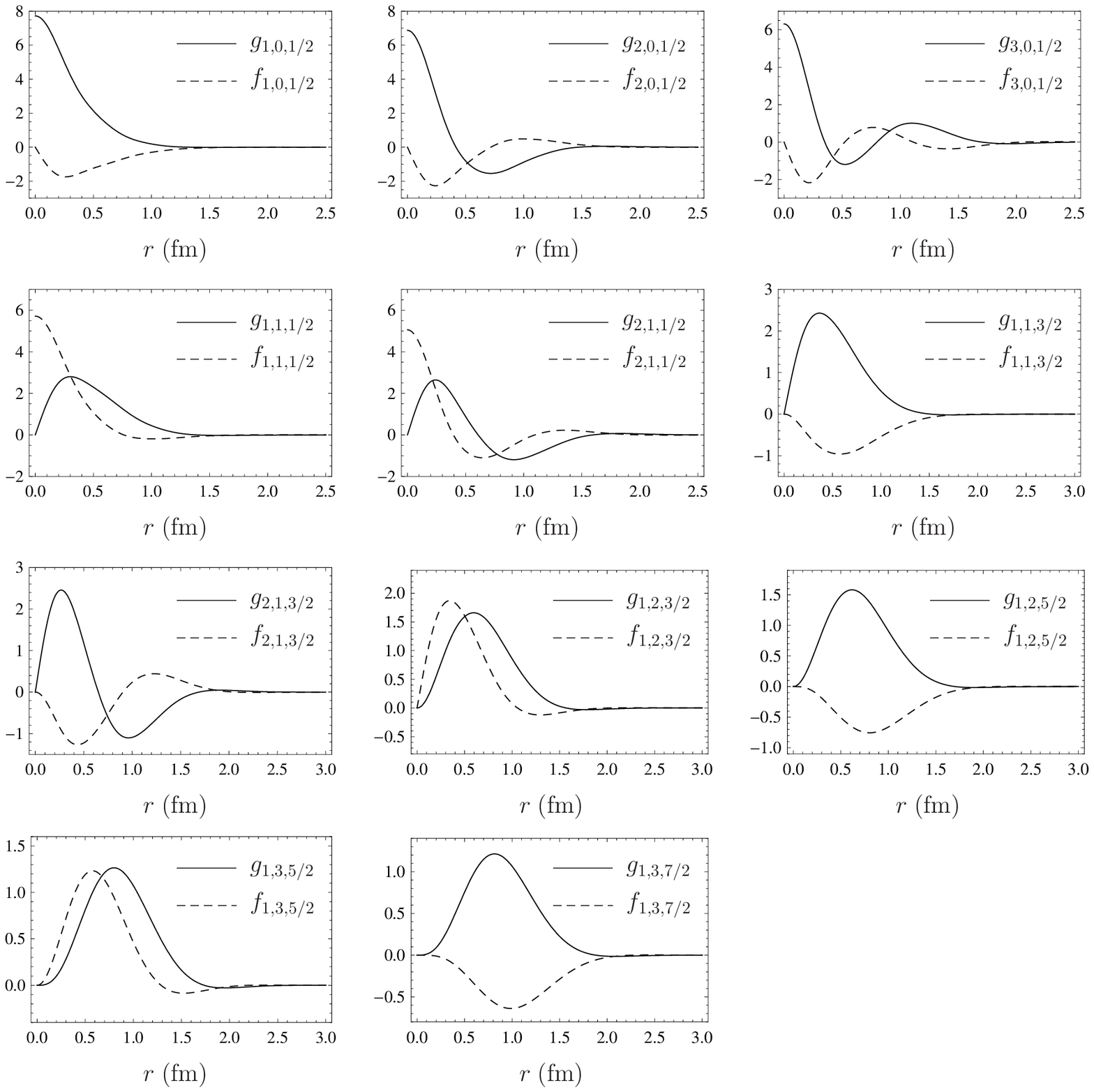}}
\caption{The radial wave functions $g_{n,l,j}(r)$ and $f_{n,l,j}(r)$ for $D$ meson as an example. The wave functions are defined in Eq. (\ref{e26}). They are the radial part of the solution of the eigenequation associated with $H_0$.}
\label{f1}
\end{figure*}

\section*{V Summary}
We construct a relativistic model for heavy-light quark-antiquark systems by studying the reduction of the instantaneous Bethe-Salpeter equation. The kernel we choose is based on scalar confining and vector Coulomb potentials, and it shows a Coulombic behavior at short distance and a linear confining behavior at long distance. The bound states of $D$, $D_s$, $B$, $B_s$ mesons are studied in this model. The predictions of the spectra and the assignments for the newly observed charmed mesons are presented. Our results are in  reasonable agreement with experimental measurements.
 The spectral structure is obtained in the relativistic model with merely three potential parameters. We find that the parameter $b$ in the confinement potential depends on the masses of the quarks in the framework of Bethe-Salpeter equation. Theoretical deviations from experimental data mainly appear in the $D_s$ meson sector, especially for the $D_{s0}^*(2317)^0$ and $D_{s1}(2460)$ resonances. The discrepancy may be ascribed to the naive assumption of the kernel, the influence
of other kernels with different spin structure can be studied in further researches. The wave functions of each bound state can be obtained by solving the wave equation and be used in the study of $B$ and $D$ decays.


\section*{Acknowledgments} This work is supported in part by the
National Natural Science Foundation of China under contracts Nos.
11375088, 10975077, 10735080, 11125525.

\section*{Appendix A: The reduction for double-heavy systems}
It is interesting to discuss the differences of the reduction results  with and without the approximation which is mentioned in section II,
\begin{equation}
 \Lambda(\boldsymbol{p})\rightarrow 1. \label{e80}
\end{equation}
In this paper we discard the approximation and retain the original form of the Salpeter equation. Here we only consider systems with vector interaction for the sake of simplicity.

As for the reduction of the double-heavy system, the procedure of the  Foldy-Wouthuysen transformation has been generalized by Chraplyvy \cite{zvc1,zvc2} in such a way that it can be applied to the two-body problem. If we write the general two-body Hamiltonian as
\begin{equation}
H_{12}=\beta^{(1)} m_1+\beta^{(2)} m_2+\mathcal{E}\mathcal{E}+\mathcal{E}\mathcal{O}+\mathcal{O}\mathcal{E}+\mathcal{O}\mathcal{O}, \label{e92}
\end{equation}
then the transformed Hamiltonian has the form
\begin{eqnarray}
&&\tilde{H}_{12}=U_{12}^{-1}HU_{12}\nonumber\\
&&=\beta^{(1)} m_1+\beta^{(2)} m_2+\mathcal{E}\mathcal{E}+\frac{\beta^{(1)}}{2m_1}(\mathcal{O}\mathcal{E})^2
+\frac{\beta^{(2)}}{2m_2}(\mathcal{E}\mathcal{O})^2 \nonumber\\
&&-\frac{\beta^{(1)}}{8m_1^3}(\mathcal{O}\mathcal{E})^4
-\frac{\beta^{(2)}}{8m_2^3}(\mathcal{E}\mathcal{O})^4
+\frac{1}{8m_1^2}[ [ \mathcal{O}\mathcal{E},\mathcal{E}\mathcal{E} ],
\mathcal{O}\mathcal{E} ] \nonumber\\
&&+\frac{1}{8m_2^2}[ [ \mathcal{E}\mathcal{O},\mathcal{E}\mathcal{E} ],
\mathcal{E}\mathcal{O} ]+\frac{\beta^{(1)}\beta^{(2)}}{4m_1m_2}\{\{ \mathcal{O}\mathcal{E},\mathcal{O}\mathcal{O} \},\mathcal{E}\mathcal{O}\} \nonumber\\
&&+\frac{\beta^{(1)}+\beta^{(2)}}{4(m_1+m_2)}(\mathcal{O}\mathcal{O})^2+\cdots \label{e93}
\end{eqnarray}

Eq. (\ref{e110}) can be written as (the wave function in the equation is omitted):
\begin{eqnarray}
&&-\frac{h_1}{2}E-\frac{h_2}{2}E+_\Delta\!\omega_1+_\Delta\!\omega_2+\frac{1}{2}(h_1+h_2)U\frac{1}{2}(h_1+h_2)
\nonumber\\
&&=-m_1-m_2,\label{ea01}
\end{eqnarray}
where
\begin{eqnarray}
_\Delta\omega_i=\omega_i-m_i,\;i=1,2\label{ea02}
\end{eqnarray}
now we consider performing the Foldy-Wouthuysen transformation on Eq. (\ref{ea01}), the left side of the equal sign can be rewritten as:
\begin{eqnarray}
&&-\frac{E}{2}\beta^{(1)}-\frac{E}{2}\beta^{(2)}-\frac{E}{2}\left(\frac{m_1}{\omega_1}-1\right)\beta^{(1)}
-\frac{E}{2}\left(\frac{m_2}{\omega_2}-1\right)\beta^{(2)}\nonumber\\
&&-\frac{E}{2}\left(-\frac{\boldsymbol{\alpha}^{(1)}\cdot\boldsymbol{p} }{\omega_1}\right)
-\frac{E}{2}\left(-\frac{\boldsymbol{\alpha}^{(2)}\cdot\boldsymbol{p} }{\omega_2}\right)\nonumber\\
&&+_\Delta\!\omega_1+_\Delta\!\omega_2+\frac{1}{2}(h_1+h_2)U\frac{1}{2}(h_1+h_2)\nonumber\\
&&=-\frac{E}{2}\beta^{(1)}-\frac{E}{2}\beta^{(2)}+\mathcal{EE}+\mathcal{OE}+\mathcal{EO}+\mathcal{OO}, \label{ea03}
\end{eqnarray}
where $U$ is defined as
\begin{eqnarray}
&&U=V_1+V_2,\\
&&V_1=V,\\
&&V_2=-\frac{1}{2}[{\boldsymbol{\alpha}}^{(1)}\cdot{\boldsymbol{\alpha}}^
{(2)}+({\boldsymbol{\alpha}}^{(1)}\cdot\hat{\boldsymbol{r}})
({\boldsymbol{\alpha}}^{(2)}\cdot\hat{\boldsymbol{r}})]V.
\end{eqnarray}

From Eq.(\ref{ea03}), we can obtain $\mathcal{EE}$, $\mathcal{OE}$, $\mathcal{EO}$ and $\mathcal{OO}$. Inserting the four operators into Eq.(\ref{e93}) and expanding the equation to order $(1/E)^2$,
we have the transformed Hamiltonian
\begin{eqnarray}
&&H_{12}=\omega_1+\omega_2+V-\frac{1}{4m_1m_2}\{{\boldsymbol{\alpha}^{(2)}}\cdot\boldsymbol{p},
\{{\boldsymbol{\alpha}^{(1)}}\cdot\boldsymbol{p},V_2\}\}
\nonumber\\
&&\frac{1}{8m_2^2}[[{\boldsymbol{\alpha}^{(2)}}\cdot\boldsymbol{p}, V],{\boldsymbol{\alpha}^{(2)}}\cdot\boldsymbol{p}]
+\frac{1}{8m_1^2}[[{\boldsymbol{\alpha}^{(1)}}\cdot\boldsymbol{p}, V],{\boldsymbol{\alpha}^{(1)}}\cdot\boldsymbol{p}] \nonumber\\
&&-\frac{{\beta^{(1)}+\beta^{(2)}}}{4E_0}(\mathcal{O}\mathcal{O})^2.\label{e95}
\end{eqnarray}
All the terms of the transformed Hamiltonian are the same as that of the Breit interaction except for the last term. In our scheme, we have
\begin{eqnarray}
&&\mathcal{O}\mathcal{O}=\frac{1}{2}(\beta^{(1)}+\beta^{(2)})V_2\frac{1}{2}(\beta^{(1)}+\beta^{(2)}),\\ &&-\frac{{\beta^{(1)}+\beta^{(2)}}}{4E_0}(\mathcal{O}\mathcal{O})^2
\rightarrow-\frac{V_2^2}{2E_0},\label{ea5}
\end{eqnarray}
in the heavy quark limit, $E_0\approx m_1+m_2$.
While with the approximation, the result is
\begin{eqnarray}
&&\mathcal{O}\mathcal{O}=V_2,\\
&&\frac{{\beta^{(1)}+\beta^{(2)}}}{4(m_1+m_2)}(\mathcal{O}\mathcal{O})^2
\rightarrow\frac{V_2^2}{2(m_1+m_2)},\label{ea6}
\end{eqnarray}
Eq.(\ref{ea5}) differs from Eq.(\ref{ea6}) by a minus sign.
In conclusion, for double-heavy system we can take the approximation and have a simpler form of wave equation if we disregard the last term in Eq. (\ref{e95}).


\end{document}